\documentclass[10pt]{article}
\usepackage{graphicx}
\usepackage{cite}
\usepackage{subcaption}
\usepackage{url}

\def\Title#1{\begin{center} {\Large #1 } \end{center}}
\def\Author#1{\begin{center}{ \sc #1} \end{center}}
\def\Address#1{\begin{center}{ \it #1} \end{center}}

\newcommand\pubblock{\rightline{\begin{tabular}{l} Proceedings of the Fifth Annual LHCP\\ \pubnumber\\
         \pubdate  \end{tabular}}}

\newenvironment{Abstract}{\begin{quotation} \begin{center} 
             \large ABSTRACT \end{center}\bigskip 
      \begin{center}\begin{large}}{\end{large}\end{center} \end{quotation}}

\newenvironment{Presented}{\begin{quotation} \begin{center} 
             PRESENTED AT\end{center}\bigskip 
      \begin{center}\begin{large}}{\end{large}\end{center} \end{quotation}}

\def\Acknowledgements{\bigskip  \bigskip \begin{center} \begin{large}
             \bf ACKNOWLEDGEMENTS \end{large}\end{center}}
\def\be{\begin{align}}
\def\ee{\end{align}}
\def\bea{\begin{align}}
\def\eea{\end{align}}

\newcommand{\secdec}{{\textsc{SecDec}}}

\newcommand{\gosam}{{\textsc{GoSam}}}

\newcommand\TwoFigBottom{-2}

\def\beq{\begin{equation}}
\def\eeq#1{\label{#1}\end{equation}}
\def\eeqn{\end{equation}}

\def\beqa{\begin{eqnarray}}
\def\eeqa#1{\label{#1}\end{eqnarray}}
\def\eeqan{\end{eqnarray}}

\let\bar=\overbar

\def\Dslash{\not{\hbox{\kern-4pt $D$}}}
\def\dslash{\not{\hbox{\kern-2pt $\del$}}}

\def\ee{e^+e^-}

\def\msb{{\bar{\ssstyle M \kern -1pt S}}}

\usepackage{color}

\newcommand\pubnumber{MPP-2017-222}
\newcommand\pubdate{\today}

\def\affiliation{
Max Planck Institute for Physics, F\"ohringer Ring 6, 80805 Munich, Germany
}

\begin{document}

% large size for the first page
\large
\begin{titlepage}
\pubblock

%% Change the title, name, abstract
%% Title 
\vfill
\Title{QCD calculations for the LHC: status and prospects}
\vfill

%  if you need to add the support use this, fill the \support definition above. 
%   \Author{ FIRSTNAME LASTNAME \support }
\Author{ Gudrun Heinrich }
\Address{\affiliation}
\vfill
\begin{Abstract}
We briefly review the status of high precision QCD predictions
available for LHC processes, focusing on corrections beyond NLO and
ways to make the latter available in a convenient format.
As a phenomenological example we discuss the two-loop corrections to
Higgs boson pair production in gluon fusion.
 \end{Abstract}
\vfill

% DO NOT CHANGE 
\begin{Presented}
The Fifth Annual Conference\\
 on Large Hadron Collider Physics \\
Shanghai Jiao Tong University, Shanghai, China\\ 
May 15-20, 2017
\end{Presented}
\vfill
\end{titlepage}
\def\thefootnote{\fnsymbol{footnote}}
\setcounter{footnote}{0}
%

% normal size for the rest
\normalsize 

%% Your paper should be entered below. 

\section{Introduction}

The LHC is collecting an impressive amount of data, entering a phase 
where many important measurements are limited by systematics rather than statistics.
As signs of physics beyond the Standard Model may hide in small deviations from the expected result,  
it is of great importance to have the Standrad Model predictions well under control.
While next-to-leading order (NLO) predictions combined with parton showering are the state of the art, 
widely used by the experimental collaborations, the number of processes where an NLO description is not sufficient is increasing
as the measurements gain in precision, stimulating remarkable progress in the theory community. 

There are several ways to improve the precision of SM predictions.
The most obvious one is to increase the order of the perturbative series in the strong coupling $\alpha_s$.
However, resumming large logarithms in certain kinematic regions and/or including electroweak corrections can sometimes be more important, 
as well as including quark mass effects. In addition, there are parametric uncertainties, for example due to limited precision on the value for 
$\alpha_s, M_W$ or $m_{\rm{top}}$, as well as uncertainties related to the parton distribution functions (PDFs), and non-perturbative effects.
In the following, a brief overview of some of these developments will be given.

\section{Progress in perturbative QCD calculations}

%\subsection{Status of calculations beyond NLO QCD}

\subsection{Building blocks of QCD corrections up to NNLO}

As NLO QCD calculations matched to a parton shower are the state of the art, we will focus on developments beyond NLO QCD.
At next-to-next-to leading order (NNLO), one can distinguish three basic contributions to the scattering amplitude:
The two-loop virtual part ${\cal A}^{\rm{VV}}$, the one-loop amplitude with one extra parton (compared to the Born configuration) that can become unresolved (soft or collinear)  ${\cal A}^{\rm{RV}}$, 
and the tree-level amplitude ${\cal A}^{\rm{RR}}$, where two extra partons can become unresolved. 
For loop-induced processes, where the leading order already proceeds via a loop (e.g. Higgs boson production in gluon fusion), the counting of the loops is of course shifted, such that the NLO virtual amplitude already involves two-loop diagrams.
If a QCD parton becomes unresolved, this entails infrared singularities which need to be isolated before any numerical integration over the phase space can be attempted. 

\subsection{NNLO double real radiation}

The isolation of the infrared singularities from ${\cal A}^{\rm{RR}}$ was for a long time a bottleneck which hampered progress in the construction of fully differential 
Monte Carlo programs for NNLO predictions. This situation however changed drastically in the last few years, 
and partly can be traced back to the fact that insights about the universal infrared behaviour of QCD, partly gained from resummation or Soft-Collinear Effective Theory, 
were used conveniently to isolate the singular regions.
%Table \ref{tab:realsub} shows some of the methods  for the isolation of infrared divergent real radiation at NNLO.

\begin{center}
\begin{table}[htb]
\caption{\label{tab:realsub}Methods for the isolation of IR divergent real
  radiation at NNLO.}
%\footnotesize\rm
\centering
\begin{tabular}{@{}*{7}{l}}
\hline
method &  analytic integr. of &type/restrictions\\
       &subtraction terms& \\
\hline
antenna subtraction~\cite{GehrmannDeRidder:2005cm}& yes&subtraction\\
$q_T$-subtraction~\cite{Catani:2009sm}&yes &slicing; colourless final states\\
N-jettiness~\cite{Boughezal:2015dva,Gaunt:2015pea}& yes&slicing &\\
sector-improved residue subtraction\cite{Heinrich:2002rc,Czakon:2010td,Boughezal:2011jf,Czakon:2014oma}&no&subtraction&\\
nested subtraction~\cite{Caola:2017dug}&no&subtraction\\
colourful subtraction~\cite{Somogyi:2006da,DelDuca:2016csb}&partly&subtraction;
colourless initial states\\
projection to Born~\cite{Cacciari:2015jma} &yes&subtraction\\
\hline
\end{tabular}
\end{table}
\end{center}
The method of $q_T$-subtraction~\cite{Catani:2009sm} or N-jettiness~\cite{Boughezal:2015dva,Gaunt:2015pea} has been employed to obtain NNLO results for LHC processes involving colourless final states or at most one jet~\cite{Catani:2011qz,Ferrera:2013yga,Grazzini:2013bna,Ferrera:2014lca,Gehrmann:2014fva,Cascioli:2014yka,Grazzini:2015nwa,Campbell:2016jau,Grazzini:2016swo,Grazzini:2015hta,Boughezal:2015aha,Boughezal:2015dva,Boughezal:2015dra,Boughezal:2015ded,Caola:2015psa,Caola:2015rqy,Grazzini:2016ctr,Boughezal:2016wmq,Campbell:2016ivq,Grazzini:2017ckn,Li:2017lbf}. 
The proceeses H+jet~\cite{Chen:2016zka}, Z+jet~\cite{Gehrmann-DeRidder:2016jns} and di-jets~\cite{Currie:2017eqf} at NNLO, as well as single jet inclusive and di-jet production in DIS~\cite{Currie:2016ytq,Currie:2017tpe} 
have been calculated based on antenna subtraction~\cite{GehrmannDeRidder:2005cm}.
Top quark pair production at NNLO~\cite{Czakon:2013goa,Czakon:2015owf} has been calculated based on sector-improved residue subtraction, 
a variant of the latter also has been used for H+jet~\cite{Boughezal:2015dra}. 
For H+2\,jets in vector boson fusion~\cite{Cacciari:2015jma} the ``projection to Born" method has been used.
A summary of schemes to treat unresolved real radiation at NNLO is given in Table~\ref{tab:realsub}.

%loop induced processes: available 2-loop corrections
\subsection{Loop integrals and two-loop amplitudes}

At two loops, many remarkable achievements can be reported,
and an impressive number of differential NNLO results for $2\to 2$ processes became available recently. 
The results for the production of two vector bosons rely on analytic calculations of two-loop four-point amplitudes with two massive/off-shell legs,
which were the subject of groundbreaking analytical work~\cite{Gehrmann:2013cxs,Henn:2014lfa,Gehrmann:2014bfa,Caola:2014lpa,Papadopoulos:2014hla,Gehrmann:2015ora,vonManteuffel:2015msa,Caola:2015ila}.

The integrals entering top quark pair production at NNLO~\cite{Czakon:2013goa} 
have been calculated numerically~\cite{Baernreuther:2013caa}, analytic results are partially available~\cite{Bonciani:2013ywa,Abelof:2015lna}. 
The NNLO corrections to top quark decay have been calculated in Refs.~\cite{Gao:2017goi,Gao:2012ja,Brucherseifer:2013iv}.
Classes of integrals with one additional mass scale appearing in the
propagators also have been calculated in the context of massive Bhabha
scattering~\cite{Henn:2013woa}, electron-muon scattering 
(with $m_e=0,m_{\mu}\not =0$)~\cite{Mastrolia:2017pfy}, and the mixed QCD-EW
corrections to the Drell-Yan process~\cite{Bonciani:2016ypc,vonManteuffel:2017myy}.

The analytic calculation of two-loop four-point  integrals with {\em both} massive propagators {\em and} massive/off-shell external legs is currently 
one of the most vibrant topics in the field of precision calculations. 
Such integrals are needed for example for the virtual NLO corrections to Higgs+jet or Higgs boson pair production in gluon fusion.
Note that the NLO corrections to these processes involve two loops, as the leading order already proceeds via a loop.
The analytic calculation of the two-loop integrals entering Higgs+jet and di-Higgs involves a new level of complexity due to the fact that the 
results contain new function classes, involving elliptic integrals, which complicate the calculation in various respects.
Nonetheless, results for the planar case have been achieved~\cite{Bonciani:2016qxi,Primo:2016ebd}. 
In addition, the top-bottom interference effects in Higgs plus jet production have been calculated~\cite{Lindert:2017pky}, 
based on the amplitudes calculated in Refs.~\cite{Melnikov:2016qoc,Melnikov:2017pgf}.
Results for planar two-loop five-point integrals~\cite{Papadopoulos:2015jft,Gehrmann:2015bfy} and certain 
helicity amplitudes~\cite{Badger:2013gxa,Badger:2015lda,Gehrmann:2015bfy,Dunbar:2016aux} also became available recently, as well as the 
two-loop six gluon all plus helicity amplitude\cite{Dunbar:2016gjb}.
A landmark in what concerns two-loop results based on a numerical unitarity method is the calculation of the full two-loop 4-gluon amplitudes presented in Ref.~\cite{Abreu:2017xsl}.

At the multi-loop front, among the most remarkable recent achievements are the five-loop QCD beta-function~\cite{Baikov:2016tgj,Luthe:2017ttg,Herzog:2017ohr,Chetyrkin:2017bjc} 
and four-loop contributions to the cusp anomalous dimension and 
N$^3$LO splitting
functions~\cite{Davies:2016jie,vonManteuffel:2016xki,Lee:2016ixa,Ruijl:2017eht,Lee:2017mip,Moch:2017uml}, 
three-loop corrections to the heavy flavour Wilson coefficients in DIS
with two different masses~\cite{Ablinger:2017err}, 
new high precision calculations of the four-loop contribution to the
electron g-2 in QED\cite{Laporta:2017okg,Marquard:2017iib}, and the N$^3$LO calculations for Higgs boson production in gluon fusion~\cite{Anastasiou:2015ema,Anastasiou:2016cez,Dulat:2017prg} and in vector boson fusion~\cite{Dreyer:2016oyx}.

\subsection{Ntuples, grids and the strong coupling}

The runtimes of NNLO programs (as well as NLO programs for multi-particle final states) which are capable of producing fully
differential results are typically rather large, even when a cluster is used for parallelized computations.
Therefore various frameworks have been developed which allow to perform scale- and PDF variations or 
the evaluation of the matrix elements at different $\alpha_s$ values without re-running the full code.
The {\tt fastNLO}~\cite{Kluge:2006xs,Stober:2015nlg} and {\tt Applgrid}~\cite{Carli:2010rw} frameworks, based on grids and an interpolation framework, 
have been developed for this purpose. 
Another possibility is to store all the relevant information needed for scale- and PDF variations in {\tt Ntuples}~\cite{Bern:2013zja,Heinrich:2016jad}. 
This is a rather storage-intensive approach, which however does not require any interpolation.

{\tt FastNLO} tables with NNLO QCD top-quark pair differential distributions have been presented in Ref.~\cite{Czakon:2017dip} and 
have proven very valuable for PDF extractions~\cite{Czakon:2016olj}. 
A recent  determination of the strong coupling constant from 
H1 data~\cite{Andreev:2017vxu}, based on an NNLO
calculation~\cite{Currie:2016ytq,Currie:2017tpe}, is another prominent
example of a successful application of the recently developed {\tt
  fastNLO/Applgrid} interface to NNLO codes~\cite{fastnloURL}. 
For other examples  see also Refs.~\cite{DelDebbio:2013kxa,Bertone:2014zva,Gao:2017kkx}.

Recent measurements of $\alpha_s$ based on LHC data can be found in Refs.~\cite{CMS:2014mna,Aaboud:2017fml}, and 
a determination of $\alpha_s$ solely based on the total cross section for  top quark pair production has been presented in Ref.~\cite{Klijnsma:2017eqp}.
The developments which led to the latest world average are summarized in Ref.~\cite{Bethke:2017uli}.

\subsection{Electroweak corrections}

As the measurements at the LHC enter the percent level precision era, electroweak (EW) corrections become increasingly important.
The need for automated tools to calculate NLO EW (and mixed QCD-EW) corrections has triggered impressive developments in the NLO community, 
see e.g. Refs.~\cite{Chiesa:2015mya,Biedermann:2017yoi} for a review. 
Recent results on NNLO QCD combined with NLO EW results for $t\bar{t}$ production have been presented in Ref.~\cite{Czakon:2017wor}. 
Other very recent achievements include the complete NLO corrections to $W^+W^+$ production and its backgrounds~\cite{Biedermann:2017bss},
di-photon+jets~\cite{Chiesa:2017gqx}, vector boson+jets\cite{Kallweit:2014xda,Lindert:2017olm}, di-bosons~\cite{Biedermann:2016yvs,Biedermann:2016guo,Biedermann:2016lvg,Kallweit:2017khh,Biedermann:2017oae},
vector boson scattering~\cite{Biedermann:2016yds}, $Ht\bar{t}$~\cite{Denner:2016wet}, $W^+W^-b\bar{b}$~\cite{Denner:2016jyo}, dijet~\cite{Frederix:2016ost}.
For more details, we refer to the contribution of J.~Lindert in these proceedings.

\subsection{N(N)LO+Parton shower}

To review the field of NLO matching to parton showers and its extension to NNLO is beyond the scope of this writeup. 
Groundwork for the NNLO+PS matching has been laid e.g. in Refs.~\cite{Lonnblad:2012ng,Hamilton:2013fea,Alioli:2013hqa,Hoche:2014dla,Hoche:2017iem}
and we will certainly see further developments in this direction.

\subsection{Jets}

Jet physics is a vast and vibrant field, which has seen very rapid progress in the past years, 
mainly due to the development of jet substructure and machine learning techniques.
For a review we refer to Ref.~\cite{Larkoski:2017jix} and the article of C.~Biino in these proceedings.
%At this point we only would like to mention that the CMS open data project was very useful to allow theorists to perform jet studies~\cite{Tripathee:2017ybi}.

%\section{Phenomenological results}

\section{Higgs boson pair production in gluon fusion}

Higgs boson pair production in gluon fusion is one of the prime processes where physics beyond the Standard Model could manifest itself. 
The leading order for this process proceeds via top quark loops, where delicate cancellations occur between  box diagrams 
and triangle diagrams, the latter involving the Higgs boson self-coupling $\lambda$.
%if $\lambda$ has its Standard Model value. 
The NLO calculation for this process with the full top quark mass dependence is complicated by the occurrence of two-loop four-point integrals involving both $m_H$ and $m_t$, which so far could not be calculated analytically. 
However, a numerical calculation, based on the program \secdec~\cite{Borowka:2015mxa,Borowka:2017idc}, 
has been presented recently~\cite{Borowka:2016ehy,Borowka:2016ypz}.
It revealed that the full top quark mass dependence reduces the cross section by about 14\% as compared to the Born-improved 
NLO HEFT approximation. 
The latter is based on ``Higgs Effective Field Theory'',  where the $m_t\to \infty$ limit has been taken
%such that the top quark loops are shrunk to a point, 
to calculate the NLO corrections, and then the NLO result is rescaled by the Born amplitude in the full theory.

\begin{figure}[htb]
\centering
\begin{subfigure}{0.49\textwidth}
\includegraphics[width=\textwidth]{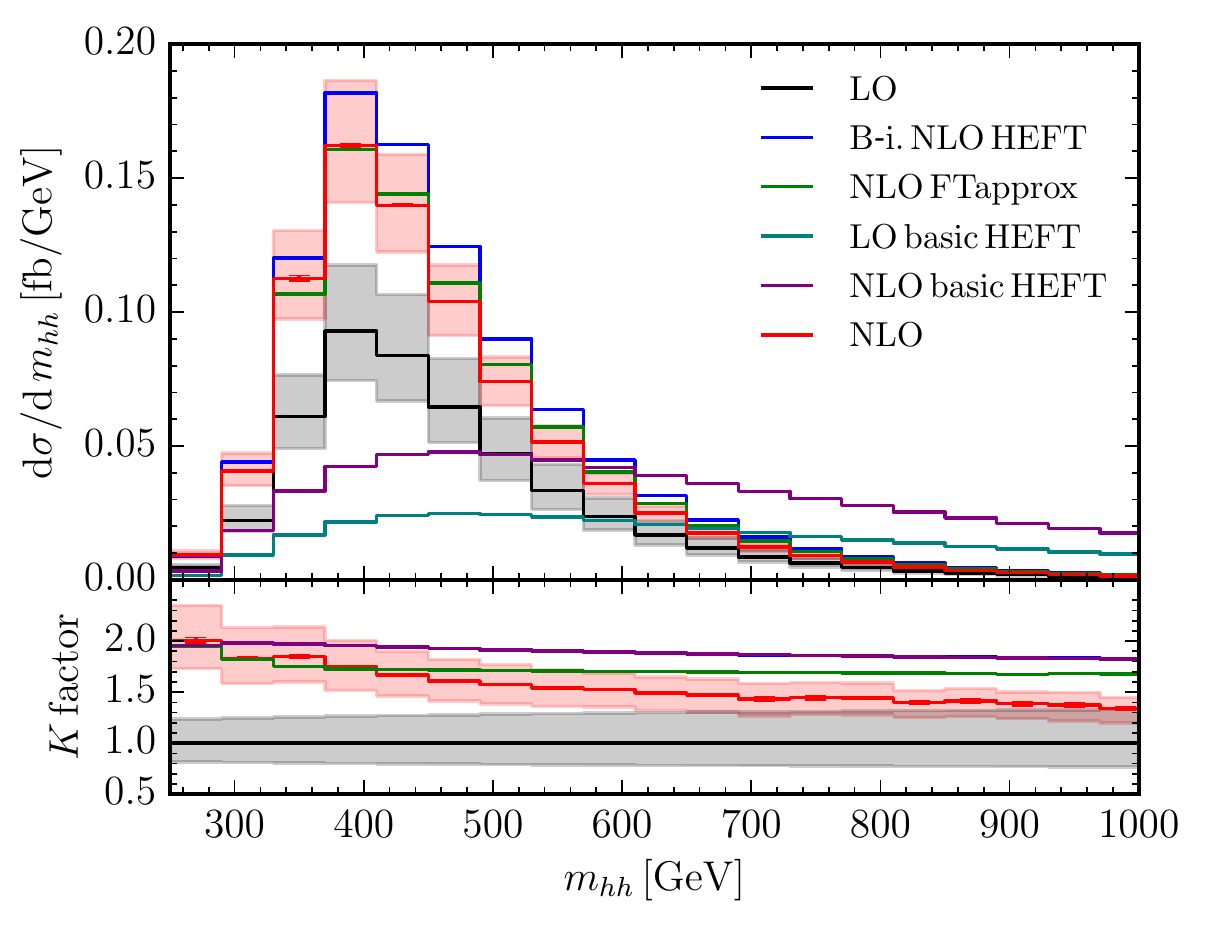}
\vspace{\TwoFigBottom em}
\caption{\label{fig:mhh}}
\end{subfigure}
\hfill
\begin{subfigure}{0.495\textwidth}
\includegraphics[width=\textwidth]{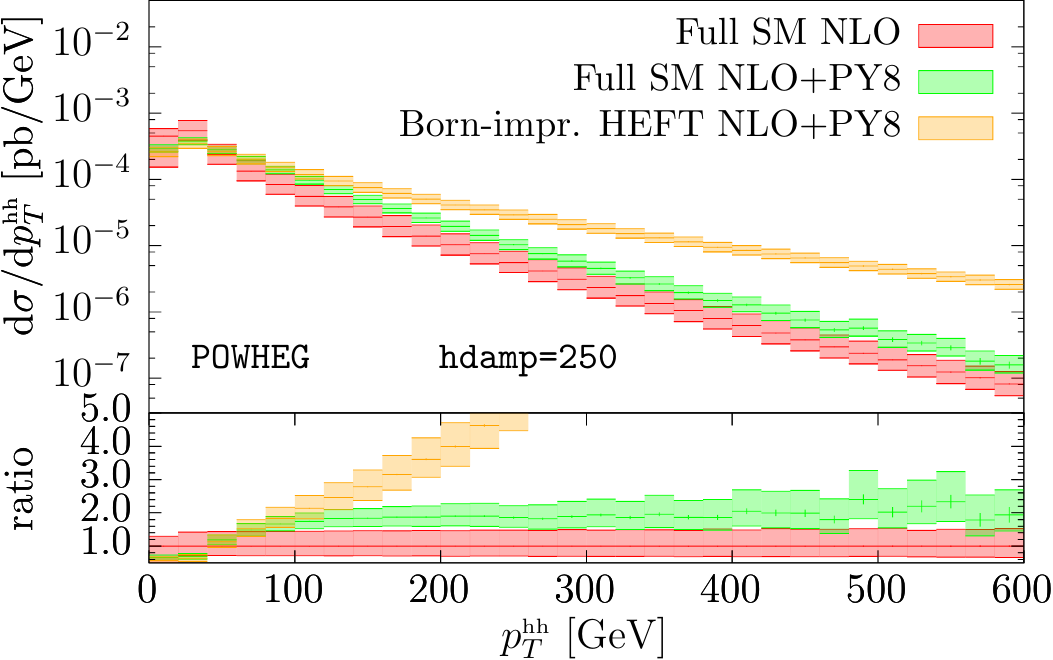}
%\vspace{\TwoFigBottom em}
\caption{\label{fig:pthh}}
\end{subfigure}
\caption{(a) Fixed order predictions for the Higgs boson pair invariant mass distribution, comparing the full result to various approximations, 
(b) NLO+parton shower results for the transverse momentum distribution of the Higgs boson pair. 
Both predictions are for the LHC at $\sqrt{s}=14$\,TeV, using the PDF4LHC15 parton distribution functions~\cite{Butterworth:2015oua}.
}
\end{figure}

The exact top quark mass dependence alters the Higgs boson pair invariant mass distribution significantly in the high $m_{hh}$ region, 
where the top quark loops can be resolved, see Fig.~\ref{fig:mhh}. The FT$_{\mathrm{approx}}$~\cite{Maltoni:2014eza} result is obtained by keeping the full top quark mass
dependence in the real radiation, while the virtual part is calculated
in the Born-improved HEFT approximation.

The numerical results for the two-loop amplitude have been implemented in a two-dimensional grid depending on the Mandelstam invariants $s$ and $t$, 
together with an interpolation procedure. This allows to combine the results with a parton shower~\cite{Heinrich:2017kxx} and to
make the results publicly available~\cite{powheg, HHgrid}, which
indeed already has been useful to validate other calculations~\cite{Grober:2017uho}.
The effect of the parton shower is moderate compared to the top quark mass effects, 
but still rather large for observables like the transverse momentum of the Higgs boson pair, $p_{T}^{hh}$, 
where NLO is the first non-trivial order to describe the tail of the distribution. 
In Fig.~\ref{fig:pthh} we show the $p_{T}^{hh}$ distribution, 
comparing fixed order results to results where a parton shower has been matched within the 
{\tt POWHEG}~\cite{Frixione:2007vw,powheg} plus {\tt Pythia8}~\cite{Sjostrand:2014zea}
framework.

%differential NNLO HEFT~\cite{deFlorian:2016uhr}

%%%%%%%%%%%%%%%%%%%%%%%%%%%%%%%%%%%%%%%%%%%%%%%%%%%%%%%%%%%%%%%%%%%%%%%%%
%%
%%   use this format to include an .eps figure into your paper
%%
%%%%%%%%%%%%%%%%%%%%%%%%%%%%%%%%%%%%%%%%%%%%%%%%%%%%%%%%%%%%%%%%%%%%%%%%%%%

\section{Summary}
%\section{Conclusions}
Precision calculations are of utmost importance in the current and
planned LHC phases, as well as at future colliders, as the
measurements will be precise enough to point to deviations from the
Standard Model at the percent level.
The past years have seen major advances in calculational techniques, 
both on the side of (multi-)loop amplitudes as well as on the real
radiation side. These developments led to an impressive increase of
fully differential NNLO predictions being available for $2\to 2$
processes, and some N$^3$LO predictions for $2\to 1$ processes.

In view of this rapid progress, it is necessary to develop efficient
ways to make these results available to a wider community, 
e.g. for PDF fits, determinations of the strong coupling constant and
inclusion into the experimental software. Ntuples or grids in various
forms may offer a solution and already have been employed successfully.
Finally, an example from Higgs boson pair production in gluon fusion
is given, where numerical results for the two-loop amplitude 
%for which an analytical result is not known, 
were encoded in a grid framework that allowed the inclusion of the
results in  parton shower Monte Carlo programs. 

It is encouraging to see that the community -- both experiment and
theory -- is rapidly moving towards a description of the data 
where the previous state of the art -- mostly NLO QCD
predictions --  is superseded in precision by various aspects, 
such as higher orders, resummation, 
electroweak corrections and quark mass effects.

%\clearpage
%%  
\Acknowledgements
I am grateful to my collaborators from the \secdec{} and \gosam{} collaborations for all the fruitful work and discussions.
I also would like to thank the organisers of the LHCP2017 conference.
This research was supported in part by the 
Research Executive Agency (REA) of the European Union under the Grant Agreement
PITN-GA2012316704 (HiggsTools).

%\begin{thebibliography}{99}

\bibliographystyle{JHEP}
\bibliography{refs_LHCP2017}

%%
%%  bibliographic items can be constructed using the LaTeX format in SPIRES:
%%    see    http://www.slac.stanford.edu/spires/hep/latex.html
%%  SPIRES will also supply the CITATION line information; please include it.
%%

%\end{thebibliography}

\end{document}